\documentclass[aps,prl,twocolumn,showpacs,showkeys,superscriptaddress]{revtex4}
\usepackage{graphicx,amsmath}

\newcommand{\be}{\begin{equation}}
\newcommand{\ee}{\end{equation}}
\newcommand{\ba}{\begin{align}}
\newcommand{\bea}{\begin{eqnarray}}
\newcommand{\eea}{\end{eqnarray}}
\newcommand{\ud}{\mathrm{d}}
\newcommand{\scH}{\mathcal{H}}
\newcommand{\vecr}{\mathbf{r}}
\newcommand{\vq}{\mathbf{q}}
\newcommand{\lan}{\left\langle}
\newcommand{\ran}{\right\rangle}

\begin{document}

\title{Effect of weak disorder on the ground state of uniaxial dipolar\\ spin systems in the upper critical dimension}

\author{A.~V.~Klopper}
\email[]{avk@cyllene.uwa.edu.au}
\affiliation{School of Physics, University of Western Australia, Perth, Australia}
\affiliation{Leibniz Institute for Solid State and Materials Research Dresden, Dresden, Germany}
\author{U.~K.~R\"o\ss ler}
\affiliation{Leibniz Institute for Solid State and Materials Research Dresden, Dresden, Germany}
\author{R.~L.~Stamps}
\affiliation{School of Physics, University of Western Australia, Perth, Australia}

\date{\today}

\begin{abstract}
Extensive Monte Carlo simulations are used to investigate the stability of the ferromagnetic ground state in three-dimensional systems of Ising dipoles with added quenched disorder. These systems model the collective ferromagnetic order observed in various systems with dipolar long-range interactions. The uniaxial dipolar spins are arranged on a face-centred cubic lattice with periodic boundary conditions. Finite-size scaling relations for the pure dipolar ferromagnetic system are derived by a renormalisation group calculation. These functions include logarithmic corrections to the expected mean field behaviour since the system is in its upper critical dimension. Scaled data confirm the validity of the finite-size scaling description and results are compared with subsequent analysis of weakly disordered systems. A disorder-temperature phase diagram displays the preservation of the ferromagnetic ground state with the addition of small amounts of disorder, suggesting the irrelevance of weak disorder in these systems.
\end{abstract}



\maketitle

\section{Introduction\label{I}}
\begin{figure}
\includegraphics[height=50mm]{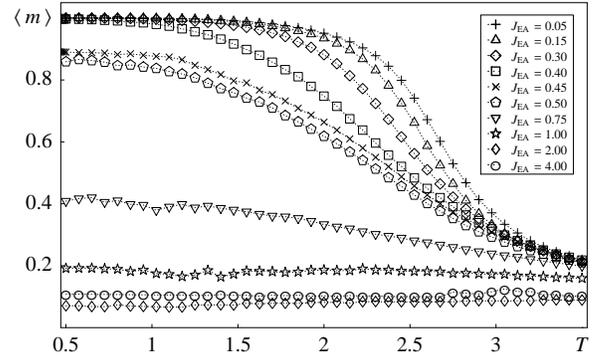}
\caption{Temperature dependence of magnetisation measurements for a FCC dipolar system with $N=108$ spins for varying amounts of disorder, controlled by the coupling constant, $J_{EA}$, which describes the strength of the randomly-signed short-range exchange.\label{Fig1-Disorder}}
\end{figure}

Certain closely-packed configurations of uniaxial dipoles can undergo a transition to a ferromagnetic ground state even in the absence of a direct exchange interaction \cite{lt46}. Dense arrangements of ferromagnetic monodomain particles and some molecular magnetic crystals are exemplary cases of this phenomenon \cite{mmlfkacdj03}. However, the existence of this dipolar ferromagnetic (DFM) ground state is highly dependent on the arrangement of spins, due to the directional dependence of the dipole interaction. Dipoles distributed on face-centred and body-centred cubic lattices \cite{bz91, bz93} are known to exhibit the transition to ferromagnetic ordering and it is straightforward (albeit computationally expensive) to simulate the transition using the Monte Carlo method. Simulation results show a tendency to order at a well-defined temperature. Such simulation models also exhibit rapid deterioration of the DFM ground state with the introduction of disorder, as shown in Fig.~\ref{Fig1-Disorder}. In this example, quenched disorder is added to a closely-packed system of Ising dipoles in the form of randomly-signed exchange between nearest neighbours, as in the short-range Edwards-Anderson (EA) spin glass \cite{ea75}. Aside from the limit in which these additional interactions dominate, the behaviour leading to the destruction of the ferromagnetic order is unknown.

Harris \cite{ab-h74} proposed a general criterion for changes to the behaviour of pure systems with the introduction of randomness. A new type of critical behaviour is predicted, if the transition observed in the pure case is sharp in a specific sense. The transition must display a power-law divergence of the specific heat (with exponent $\alpha>0$), as opposed to a singular thermodynamic behaviour (with $\alpha<0$), where the transition is signalled only by a cusp in specific heat. Chayes \emph{et al.} \cite{ccfs86} examine a host of systems for which a positive specific heat exponent in the pure system corresponds to a negative exponent in the random equivalent. Indeed, if the converse is true, and the pure system has a negative specific heat exponent, then the addition of disorder simply leads to irrelevant corrections to the scaling behaviour \cite{pv00}.

Three-dimensional Ising dipolar systems have been shown to behave in a way characteristic of a mean field description \cite{lk69, a-a73, bzj76}, whereas those comprising vector dipolar spins typically resemble systems interacting via short-range exchange \cite{rmfk95}. This discrepancy stems from the fact that the upper critical dimension (defining the lower limit for which fluctuations can be neglected) is three in the former case and four in the latter. The Ising dipolar system therefore represents a unique case for which predictions based on exact solutions from mean field theory at marginal dimensionality can be tested experimentally \cite{reyrab90}. Furthermore, it presents an ideal testing ground for the extent to which the interaction range influences critical behaviour \cite{lb97}. The marginality of the system manifests itself in the appearance of logarithmic corrections to mean field theory which produce a logarithmic divergence in the specific heat, implying a sharp transition, consistent with experimental findings for three-dimensional systems of Ising dipoles \cite{akg75}.

This investigation examines the effect of disorder on a three-dimensional system of Ising dipoles by means of Monte Carlo simulations. Finite-size scaling is employed to locate the temperature of the expected phase transition in the pure system and characterise its critical properties. Data from weakly disordered systems are scaled in a similar manner in order to gauge the relevance of the disorder. It is clear, both from the raw data and from the scaling analysis presented, that the ferromagnetic ground state cannot be detected beyond a certain disorder level.

\section{Monte Carlo simulations\label{MCS}}
The model Hamiltonian for the uniaxial dipole system with the Ising spin axis oriented in the $z$ direction \cite{af73} is,
\be\label{Hamiltonian 1}
\scH=-\frac{1}{2}J_{EA}\sum _{\vecr}\sum _{\delta }\sigma_{\vecr\mathbf{\delta}}S_{\vecr}S_{\vecr+\mathbf{\delta}
}-G\sum _{\vecr\neq \vecr^{\prime }}\frac{\partial ^{2}}{\partial
z^{2}}\frac{S_{\vecr}S_{\vecr^{\prime }}}{|\vecr-\vecr^{\prime }|^{ d-2}}.
\ee
The vector $\mathbf{\delta}$ has length $a$ and runs over $c$ nearest neighbours, coupling spins via a quenched random bond variable, $\sigma_{\vecr\mathbf{\delta}}\in\{-1,+1\}$. The magnitude of the exchange energy is denoted $J_{EA}$ and $G=\frac{1}{2}{(g \mu _{B})}^{2}$ is the strength of the dipole interaction. Spins are placed on a lattice array with face-centred cubic (FCC) structure and number density of one spin per unit cube, in the length scale of the simulated system. The dipole axes are aligned with one of the edges of the cubic lattice cells. Periodic boundary conditions are implemented by Ewald summation for a cubic simulation box \cite{kb79}. Energy, temperature and $J_{EA}$ are all measured in units of dipolar interaction energy for a pair of spins of unit length separation.

Using the Metropolis Monte Carlo method, data are obtained for systems of linear dimension $L=3$, $4$, $5$, $6$ and $7$ via a process of slow-cooling. This is executed through a series of isothermal runs, in temperature steps of 0.025 and 0.05. After an initial prologue for equilibration, several thermodynamic properties are measured, including magnetisation, $\lan m\ran$; second and fourth magnetic moments, $\lan m^{2}\ran$ and $\lan m^{4}\ran$; energy and its square, $\lan E\ran$ and $\lan E^{2}\ran$; susceptibility, $\chi=(\lan m^{2}\ran-\lan m\ran^2)/T$; specific heat, $c=(\lan E^{2}\ran-\lan E\ran^2)/T$; and the Binder cumulant, $U=1-\lan m^{4}\ran/3\lan m^{2}\ran^{2}$. Disorder is introduced in the form of quenched random interactions between nearest neighbours. These are short-range EA-type spin glass interactions of variable strength, $J_{EA}$, and disorder averages are determined from tens to hundreds of disorder realisations.

\section{Renormalisation group transformation\label{RGT}}
The critical properties of the system are determined by scaling the data to achieve size-independent collapse of the temperature dependence of all observables. The required finite-size scaling Ans\"atze are derived by examining the behaviour of the system under renormalisation group transformation. Since the test system comprises purely dipolar interactions, the random variable is omitted from the Hamiltonian in Eq.~(\ref{Hamiltonian 1}). Whilst the framework for this transformation was developed many years ago \cite{af73}, the form of the scaling functions for the marginal case of the Ising dipolar system is yet to be explicitly derived. One can formalise the `phenomenological renormalisation' \cite{mp-n75} occuring at the transition by transforming the lattice size $L$ as $L^{\prime}=L/b$.

The modified Hamiltonian is converted to one describing continuous spins via the standard $\phi^{4}$ theory in which a coefficient $u_{0}$ acts as a control parameter for the interaction \cite{dj-a84}. A source term, $h$, is included to enable functional differentiation of the free energy. The result takes the form,
\bea\label{Hamiltonian 2}
\overline{\scH}=&&-\frac{1}{2}\int_{q}U_{0}(q) \phi _{q}\phi _{-q}-h\phi_{q=0}\nonumber\\
&&-u_{0}\int_{q_{1}}\int_{q_{2}}\int_{q_{3}}\phi _{q_{1}}\phi
_{q_{2}}\phi _{q_{3}}\phi _{-q_{1}-q_{2}-q_{3}},
\eea
where the following integral convention is employed hereafter, $\int_{q} \to \int_{0\leq q\leq \Lambda }\frac{\ud^{d}\vq}{{( 2\pi )}^{d}}$ with $\Lambda \sim \pi /a$. The function in the first integral of Eq.~(\ref{Hamiltonian 2}) is defined as,
\be\label{U-0}
U_{0}( q ) =r_{0}+q^{2}-f_{0} {( q^{z}) }^{2}+g_{0} {\left(
\frac{q^{z}}{q}\right) }^{2},
\ee
where the coefficient $r_{0}\propto t=(T-T_{c})/T_{c}$ and $T_{c}$ is the critical temperature of the DFM transition. The remaining coefficients arise from an expansion of the dipolar term in the Hamiltonian in the limit of small $q$. The spin-spin correlation function for the Gaussian approximation to this model is simply the inverse of this function, $U_{0}(q)$.
 For $q=0$, the correlation function is shape-dependent and takes the form,
\be\label{Shape-Dependent Greens Function}
G_{0}(0)=(r_{0}+g_{0}(d^{-1}-(d-2)\Phi/a_{1}))^{-1},
\ee
where,
\be\label{Shape-dependent Sum}
\Phi=-\sum_{j}r_{ij}^{-d}(1-d\cos^{2}\theta_{ij}),
\ee
and $\theta_{ij}$ is the angle between $r_{ij}$ and the $z$ axis \cite{a-a73}.
 The renormalisation is undertaken by integrating out large wavevector modes and expanding the non-Gaussian part of the Hamiltonian perturbatively for small $u_{0}$ \cite{kr}. The result is rescaled according to $q=q^{\prime}/b$ and $\phi=\zeta\phi_{q^{\prime}}^{\prime}$, where $\zeta$ is chosen in order to keep the coefficient of the $q^{2}$ term constant to leading order in $u_{0}^{2}$. The renormalisation group equations can be written in differential form by taking $b=e^{\ell}$ and building up infinitesimal transformations using the group composition property.

The resulting scaling function for the free energy density takes the form,
\be\label{SF for FE 1}
f( t,h,u,L^{-1}) =b^{-d}f( t^{\prime },h^{\prime },u^{\prime },b\,L^{-1})
+g( t,h),
\ee
where $g$ represents the analytic part of the transformation and coefficients $r_{0}$ and $u_{0}$ have been identified with variables $t$ and $u$. The latter is a dangerous irrelevant variable since the free energy can become singular below the transition if it is removed. The correct finite-size scaling properties in this case are obtained by renormalising the system size to unity \cite{lb96}. A further substitution of $\phi ^{\prime }=\phi/{u^{\prime }}^{1/4}$ then gives a new scaling form for the free energy,
\be\label{SF for FE 2}
f( t^{\prime },h^{\prime },u^{\prime },1) +\overline{g}( t,h)
=\tilde{f}( \tilde{t},\tilde{h}),
\ee
\be
\tilde{t}=t^{\prime }/{u^{\prime }}^{1/2},\tilde{h}=h^{\prime
}/{u^{\prime }}^{1/4}.
\ee
The analytic part of the transformation also contributes to the singular dependence of the free energy on $t$ and this is absorbed into the function $\tilde{f}$. In order to write the scaling forms in a representation conducive to data fitting, it is convenient to absorb all constant factors into the singular function and define a fitting parameter, $v$, which controls the shift imposed on the reduced temperature variable, $t$ \cite{gh04}. Combining Eq.~(\ref{SF for FE 1}) and Eq.~(\ref{SF for FE 2}), the free energy includes logarithmic corrections and is expressed as,
\be\label{SF for FE 3}
f( t,h,u,L^{-1}) =
L^{-d}\tilde{f}(\tilde{t},\tilde{h}) +\tilde{g}( t,h),
\ee
\be
\tilde{t}= L^{y_{t}}\log^{1/6}L(t+v L^{-y_{t}}\log^{-2/3}L) , \tilde{h}=L^{y_{h}}h \log^{1/4}L,
\ee
where $y_{t}=3/2$ and $y_{h}=9/4$ define the renormalisation exponents for the mean field theory and logarithmic functions are written in terms of some unitary length scale \cite{lb97}.

\section{Finite-size scaling\label{FSS}}
The scaling forms for physical quantities follow then from simple derivatives of this function. They can be expanded in Taylor series about the finite-size limit in the critical region, given by small $t$ and finite $L$ \cite{blh95}. The functions are,
\bea\label{SF for M}
\lan  m\ran  \sim && L^{y_{h}-d}\log^{1/4}L\,\tilde{F}_{\lan m\ran}(\tilde{t},\tilde{h})\nonumber\\
\sim && L^{y_{h}-d}\log^{1/4}L \left(
a_{0}+a_{1}\tilde{t}+\frac{1}{2}a_{2}{\tilde{t}}^{2}+...\right.\nonumber\\
&& \left. +\frac{b_{1}}{\log L }+\frac{b_{2}}{\log^{2}L }+...\right),
\eea
for the magnetisation,
\bea\label{SF for M2}
\lan  m^{2}\ran \sim && L^{2y_{h}-2d}\log^{1/2}L\,\tilde{F}_{\lan m^{2}\ran}(\tilde{t},\tilde{h})\nonumber\\
\sim && L^{2y_{h}-2d}\log^{1/2}L
\left( a_{0}+a_{1}\tilde{t}+\frac{1}{2}a_{2}{\tilde{t}}^{2}+...\right.\nonumber\\
&& \left. +\frac{b_{1}}{\log
L }+\frac{b_{2}}{\log^{2}L }+...\right),
\eea
for the second order magnetic moment,
\bea\label{SF for Chi}
\chi \sim && L^{2y_{h}-d}\log^{1/2}L\,\tilde{F}_{\chi}(\tilde{t},\tilde{h})\nonumber\\
\sim && L^{2y_{h}-d}\log^{1/2}L \left( a_{0}+a_{1}\tilde{t}+\frac{1}{2}a_{2}{\tilde{t}}^{2}+...\right.\nonumber\\
&& \left. +\frac{b_{1}}{\log
L }+\frac{b_{2}}{\log^{2}L }+...\right),
\eea
for the susceptibility and,
\bea\label{SF for U}
U \sim && \tilde{F}_{U}(\tilde{t},\tilde{h})+c_{1}L^{d-2y_{h}}\nonumber\\
\sim && a_{0}+a_{1}\tilde{t}+\frac{1}{2}a_{2}{\tilde{t}}^{2}+...\nonumber\\
&& +\frac{b_{1}}{\log
L }+\frac{b_{2}}{\log^{2}L }+...+c_{1}L^{d-2y_{h}},
\eea
for the Binder cumulant. The last term in the scaling form for the Binder cumulant arises due to the field dependence of the analytic part of the free energy. This correction is not expected to affect the scaling of the other thermodynamic properties, as the logarithmic corrections will dominate their form \cite{lb97}. Data collapse is achieved by plotting scaled observables with respect to the shifted temperature variable, $\tilde{t}$, so that data from systems on all length scales fall onto a single curve.

The scaling form is only valid in the vicinity of the transition so the fitted data are restricted to this region in order to achieve meaningful collapse. The imposed restrictions are determined by pinning the set of each sample size to that of the smallest size. The subset of the smallest sample is chosen by minimising the weighted cost associated with fitting the restricted Binder cumulant data to the mean field approximation of Eq.~(\ref{SF for U}).

\begin{figure*}
\includegraphics[height=50mm]{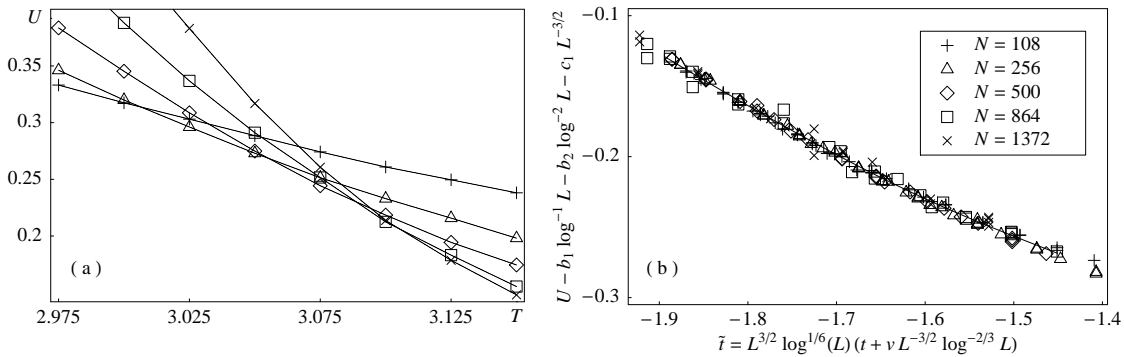}
\caption{(a) Finite-size dependence and (b) collapse of the Binder cumulant data in the critical region for the pure dipolar system.\label{Fig2-FD}}
\end{figure*}

\begin{table*}
\caption{Parameter values for fitting the pure dipolar system, quoted with $99\%$ confidence levels.\label{Tab1-Parameters}}
\begin{tabular}{|c|c c|c c c|}
\cline{2-6}
\multicolumn{1}{c|}{} & $T_{c}$ & $v$ & $b_{1}$ & $b_{2}$ & $c_{1}$\\
\hline
$\lan m\ran$ & $3.15292\pm 0.00013$ & $-1.5317\pm 0.0011$ & $0.44\pm0.10$ & $-0.36\pm0.08$ &\\
$\lan m^{2}\ran$ & $3.15292\pm 0.00005$ & $-1.14183\pm0.00042$ & $0.79\pm0.17$ & $-0.62\pm0.13$ &\\
$\chi$ & $3.152916\pm 0.000004$ & $-1.14531\pm 0.00004$ & $0.08\pm0.02$ & $-0.07\pm0.01$ &\\
$U$ & $3.152916\pm 0.000005$ & $-1.64751\pm 0.00004$ & $1.30\pm0.05$ & $-0.31\pm0.01$ & $-2.33\pm0.19$\\
\hline
\end{tabular}
\end{table*}

The ability of a given set of fit parameters to obtain acceptable data collapse can be determined via a suitably robust \textit{ad hoc} method. One scales every set of data points for all lattice sizes using the appropriate $L$-dependent logarithmic factors. Combining them in a single set, the data are then interpolated as a high-order polynomial defining an effective mean. The $L$-dependence of this function is implicit in the functional form of $\tilde{t}=\tilde{t}(T,L;T_{c},v)$.

A cost function is formed from the error-weighted sum of squares for the vertical deviation of each scaled data point from this master fit curve. Minimising this function numerically yields a set of parameters for each observable. A global value for the critical temperature, $T_{c}$, is obtained by averaging the weighted $T_{c}$ values from the initial fits and refitting all observables using the fixed average value.

\section{Results\label{R}}
In the case of the pure dipolar system, the dimensionless ratio $U$ displays scale invariance at the transition with some statistical scatter. It is clear in Fig.~\ref{Fig2-FD}(a) that there is a size-dependent systematic shift away from a single scale invariant crossing at the critical temperature. Fig.~\ref{Fig2-FD}(b) shows the collapse of these data, confirming the validity of the derived scaling functions. The critical temperature and the fit parameter $v$ are only effective in collapsing the size-dependent data sets laterally. The vertical displacement required to achieve total collapse is controlled by parameters $b_{1}$ and $b_{2}$ of the logarithmic corrections and the importance of their inclusion is vast when compared with previous analyses of Ising dipole systems \cite{xbr92}. Global fits for magnetisation and susceptibility also show excellent data collapse \cite{kr}. The correction arising from the analytic part of the free energy density transformation has little effect for all observables aside from the Binder cumulant, and this is a further indication of the importance of the logarithmic corrections in the upper critical dimension of the uniaxial dipolar system. It should be noted that the least sum of squares values corresponding to the parameters given by the global $T_{c}$ deviate by no more than $2\%$ from their original values in the initial fits.

Fitted parameter values are quoted with error estimates corresponding to $99\%$ confidence in Table~\ref{Tab1-Parameters}. The fitted values for the lateral parameters, $T_{c}$ and $v$, are calculated with much greater confidence than are those associated with vertical shift, which pertain to logarithmic corrections. This is to be expected, since the refitting process is independent of $T_{c}$, and the value chosen for $v$ is pinned to that of the critical temperature.

\begin{figure*}
\includegraphics[height=50mm]{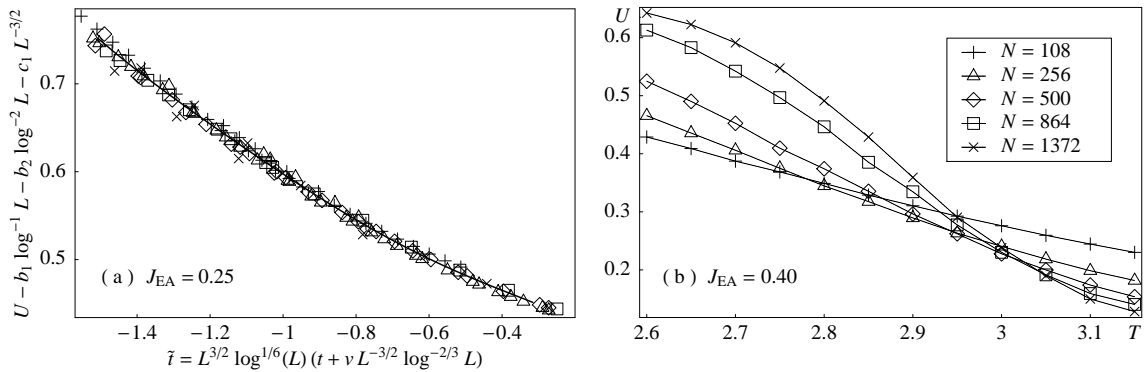}
\caption{(a) Collapse of the Binder cumulant data for the dipolar system with $J_{EA}=0.25$. (b) Temperature variation in Binder cumulant data for $J_{EA}=0.40$ suggests a ferromagnetic transition still exists.\label{Fig3-EA}}
\end{figure*}

The mean field scaling functions yield good data collapse for a certain range of weak disorder. Fits are comparable with those of the pure case up to a disorder strength of $J_{EA}\sim 0.25$. As shown in Fig.~\ref{Fig3-EA}(a), the mean field predictions are applicable at this value and the data sets collapse as they did in the pure case. Beyond this disorder strength, the simulations continue to show a transition to ferromagnetic ordering (see Fig.~\ref{Fig3-EA}(b)). However, fitting data to the mean field scaling forms proves increasingly difficult in this intermediate range and confidence levels suggest the description is unsuitable here. When disorder exceeds a level corresponding to $J_{EA}\sim 0.5$, there is no ferromagnetic transition and data cannot be scaled at all using the mean field predictions. This range coincides with the disorder level at which the ferromagnetic transition appears to deteriorate in Fig.~\ref{Fig1-Disorder}.

\section{Discussion\label{D}}
\begin{figure}[b]
\includegraphics[height=50mm]{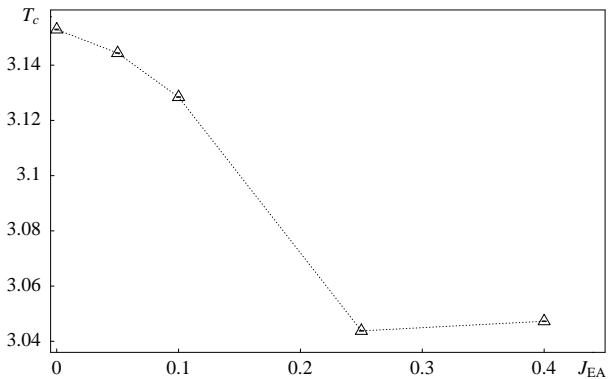}
\caption{Variation in critical temperature, $T_{c}$, with increasing disorder strength, $J_{EA}$.\label{Fig4-Tinf}}
\end{figure}

The critical temperature of the apparent ferromagnetic transition calculated by the fitting process decreases as the disorder strength is increased, as shown in Fig.~\ref{Fig4-Tinf}. The smooth change away from the value used to fit data for the pure case supports the preservation of the transition to DFM behaviour predicted by the mean field theory. Taking into account the numerical inaccuracy of the disordered system data, this smoothness is evident in the case where added disorder does not exceed $J_{EA}=0.25$, whereupon there is an abrupt change from the low disorder trend. This does not necessarily mean that systems beyond this disorder range are not of DFM character, since the mean field scaling functions can still achieve data collapse. However, discontinuities in the variation of parameter values may point towards a \textit{change} in the critical behaviour.

There are two plausible scenarios in which this change could occur. The first predicts an intermediate transition from pure DFM behaviour to disordered DFM behaviour at some finite strength $J_{EA}=J^{*}$. One might suppose this new disordered phase exists in systems with disorder strength below some upper limit, $J_{EA}=J^{\dagger}$, at which spin glass ordering sets in. An alternative picture describes a gradual crossover to a new kind of disordered critical behaviour for any finite value of $J_{EA}$. However, this scenario contradicts the results of scaling in the low disorder region to some extent. In particular, the onset of new behaviour with the introduction of disorder is not expected to be well-described by the mean field theory, as seen here. However, it is conceivable that the simulated systems are too small to detect such a subtle change.

Whilst the behaviour of the system with intermediate disorder strength remains unclear, one can be confident that the strongly disordered systems simulated here do not have a DFM ground state. The behaviour observed in this region of phase space cannot be determined within the mean field description presented. Further analysis must address a revision of the scaling functions for the thermodynamic properties of disordered uniaxial dipolar systems, and an investigation of alternate forms of ordering for the limit of strong disorder.

\begin{acknowledgments}
We acknowledge the Australian Research Council and the UWA Graduates Association for support. AVK thanks H.~Eschrig \& U.~Nitzsche for hospitality and support at IFW Dresden.
\end{acknowledgments}

\bibliography{Klopper}

\end{document}